%% file: main.tex
\documentclass{sig-alternate}

\usepackage{graphics,graphicx}
\usepackage{subfigure}
\usepackage{color}
 \usepackage{stfloats}

\usepackage{amssymb,amsmath}
\usepackage{xspace}
\usepackage{wrapfig}

\usepackage{balance}

\begin{document}

\clubpenalty=10000
\widowpenalty = 10000
\displaywidowpenalty = 10000

\toappear{}

\author{
Stephan G\"unnemann{\large${\-^{\circ\bullet}}$} \hspace{0.4cm} Hardy Kremer{\large${\-^\circ}$} \hspace{0.4cm}  Matthias Hannen{\large${\-^\circ}$} \hspace{0.4cm}
Thomas Seidl{\large${\-^\circ}$}\\[3mm]
\begin{tabular}{ccc}
 \affaddr{ {\large${\-^\circ}$}RWTH Aachen University, Germany} && {\large${\-^\bullet}$}\affaddr{Carnegie Mellon University, USA}\\
\email{\{lastname\}@cs.rwth-aachen.de} && \email{sguennem@cs.cmu.edu}\end{tabular}
}

\title{KDD-SC: Subspace Clustering Extensions for\\ Knowledge Discovery Frameworks}

\maketitle

\begin{abstract}
Analyzing high dimensional data is a challenging task. 
For these data it is known that traditional clustering algorithms fail to detect meaningful patterns. 
As a solution, subspace clustering techniques have been introduced. 
They analyze arbitrary subspace projections of the data to detect clustering structures.

In this paper, we present our subspace clustering extension for KDD frameworks, termed KDD-SC. 
In contrast to existing subspace clustering toolkits, our solution neither is a standalone product nor is it tightly coupled to a specific KDD framework.
Our extension is realized by a common codebase and easy-to-use plugins for three of the most popular KDD frameworks, namely KNIME, RapidMiner, and WEKA. 
KDD-SC extends these frameworks such that they offer a wide range of different subspace clustering functionalities. 
It provides a multitude of algorithms, data generators, evaluation measures, and visualization techniques specifically designed for subspace clustering.
These functionalities integrate seamlessly with the frameworks' existing features such that they can be flexibly combined. 
KDD-SC is publicly available on our website. 
\end{abstract}

\input{intro.tex}

\input{extension.tex}

\input{scenario.tex}

\balance

\end{document}

%% file: intro.tex
\section{Introduction}

\begin{figure}
	\centering
\includegraphics[width=0.475\textwidth]{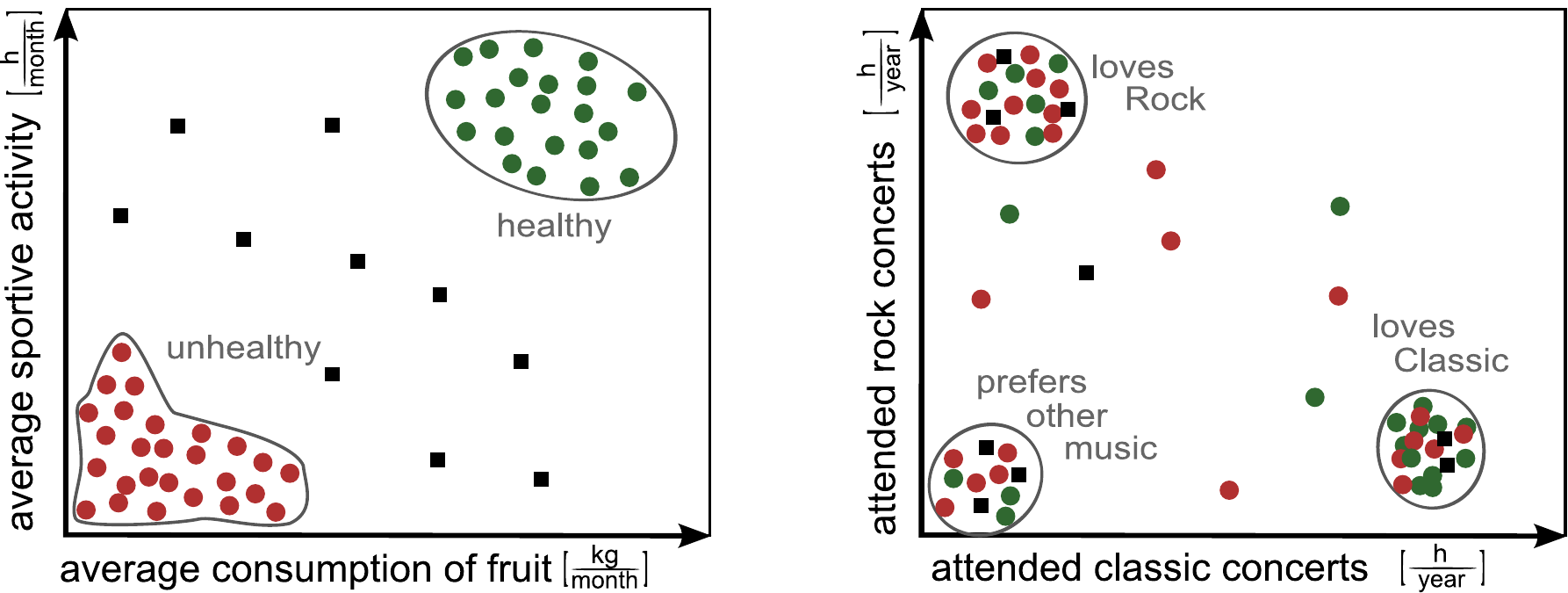}
	\caption{Subspace clusters hidden in locally relevant subspace projections}\label{fig:intro}
\end{figure}

Clustering is one of the core data mining tasks. The goal of clustering is to automatically group similar objects while separating dissimilar ones. Traditional clustering methods consider all dimensions of the dataspace to measure the similarity between objects. For today's high dimensional data, however, these full-space clustering approaches fail to detect meaningful patterns since irrelevant dimensions obfuscate the clustering structure \cite{BGR+99,DBLP:journals/tkdd/KriegelKZ09}.
Using global dimensionality reduction techniques such as principle components analysis is not sufficient to solve this problem: by definition, all objects are projected to the same lower dimensional subspace. However, as Figure~\ref{fig:intro} illustrates, each cluster might have locally relevant dimensions and objects can be part of multiple clusters in different subspaces. These effects cannot be captured by global dimensionality reduction approaches. 

To tackle this challenge, subspace clustering techniques have been introduced, aiming at detecting locally
relevant dimensions per cluster \cite{DBLP:journals/tkdd/KriegelKZ09,DBLP:journals/sigkdd/ParsonsHL04}. They analyze arbitrary subspace projections of the data to detect the hidden clusters. Typical applications for subspace clustering include gene expression analysis, customer profiling, and sensor network analysis. In each of these scenarios, subsets of the objects (e.g., genes) are similar regarding subsets of the dimensions (e.g., different experimental conditions).

\emph{Existing systems:} Today, general data mining functionality is provided to the end-user in a convenient and intuitive way by established knowledge discovery frameworks as KNIME (Konstanz Information Miner, \cite{ref:KNIME}), RapidMiner \cite{ref:rminer}, and WEKA (Waikato Environment for Knowledge Analysis, \cite{ref:WEKA}). These systems are succesfully and frequently used in research and practice. The applicability of subspace clustering, in contrast, is still limited.

So far, there are two systems that support the user in the task of subspace clustering, namely OpenSubspace \cite{DBLP:journals/pvldb/MullerGAS09} and ELKI \cite{DBLP:conf/ssdbm/AchtertKZ08}.
Both systems are milestones in the process of providing subspace clustering functionality to the end-user, but have severe limitations concerning their integration into established data mining workflows.
While ELKI, as a stand-alone java framework, does not offer any integration into existing data mining toolkits, OpenSubspace is highly coupled and its current form only applicable within the WEKA framework. 
Due to this strong coupling, it is difficult to integrate new algorithms and to (re)use already implemented subspace clustering functionality in other KDD frameworks.
Accordingly, for end-users running their established KDD workflows in other frameworks than WEKA or ELKI, the integration of subspace clustering into these workflows is a hard and time-consuming challenge.

\emph{Our contribution:} In this paper, we propose a new system for subspace clustering, which is seamlessly integrated into KNIME, RapidMiner, and WEKA.  By covering this broad spectrum of knowledge discovery frameworks, many researchers and practitioners can benefit from our system.
It is based on a common code basis across all KDD frameworks. Thus, it is possible to quickly deploy new subspace clustering methods in multiple frameworks at the same time.

 By integrating our system into these established knowledge discovery frameworks, the user can easily use subspace clustering functionality within the whole KDD process. Our methods can be combined with the existing algorithms, data transformations techniques, and visualization tools of these frameworks. 
Overall, our system offers
{\setlength{\leftmargini}{15pt}
\begin{itemize}\setlength{\itemsep}{3pt}
	\item a seamless integration of subspace clustering functionality into KNIME, RapidMiner, and WEKA. Accordingly, many researchers and practitioners can use their established KDD workflows without any loss in productivity. 
	\item a common code basis for subspace clustering algorithms, evaluation measures, and synthetic data generators. 	
	It is independent of the chosen data mining framework and realizes easy extensibility and reusability of all components. 
	\item visualization and interaction principles for subspace clustering exploiting the capabilities of the different data mining toolkits, which support the user in the interpretation of the obtained results.
\end{itemize}}

%% file: extension.tex
\section{General Architecture}

In this section we describe the general architecture and functionality of our subspace clustering extension. The usage of our extension within the different knowledge discovery frameworks is described in the Sections \ref{sec:knime}-\ref{sec:weka}.

\begin{figure}[h]
	\centering
		\includegraphics[width=0.355\textwidth]{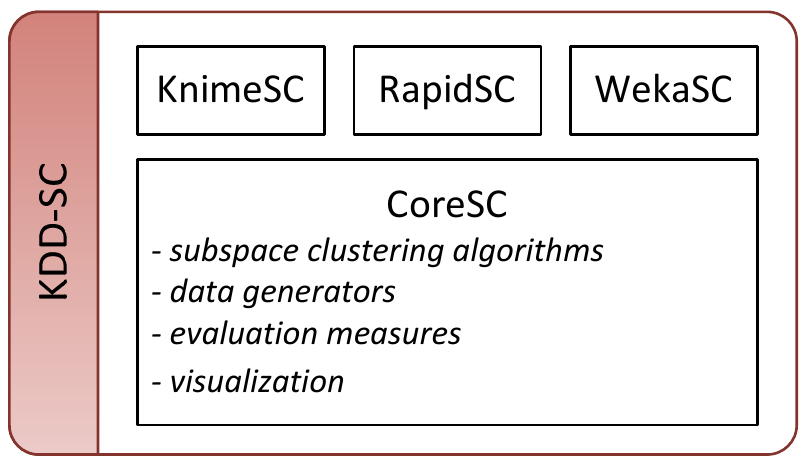}
	\caption{General Architecture of KDD-SC}\label{fig:framework}
\end{figure}

For reusability and easy portability of the developed methods, our KDD-SC framework is separated into a core package (CoreSC) and packages realizing the integration into the different KDD frameworks (KnimeSC, RapidSC, WekaSC). Figure~\ref{fig:framework} shows an overview of this design.

In the core package, the actual functionality of our system is implemented. This functionality is independent of a specific system. The core package is divided into four major components: subspace clustering algorithms, data generators, evaluation measures, and visualization tools. A detailed description of these components is provided in the following sections. 

The WekaSC user interface as well as parts  of the core package (algorithms \& evaluation measures) have been extracted from the OpenSubspace project \cite{DBLP:journals/pvldb/MullerGAS09}. 
In contrast to the original OpenSubspace, which was tightly bundled with a specific WEKA version, our redesigned WekaSC implements the WEKA plugin interface and enables easy extension by our component-based design.

In the three packages KnimeSC, RapidSC, and WekaSC we included the implementations which are necessary to realize an interaction of the knowledge discovery frameworks with the core package. Thus, these packages act as adapters between the core package and the actual system. In the KnimeSC package, for example, we implemented the node-based representation of the algorithms as required for the KNIME framework (cf. Section~\ref{sec:knime}). 

By using a common codebase, i.e.\ the CoreSC package, it is easy to integrate new subspace clustering techniques for each of the knowledge discovery frameworks. The actual subspace clustering algorithm has only to be implemented in the core package.
Additionally, one can easily support other (e.g., R) or even new data mining frameworks by simply providing a new adapter package.  

\vspace{0.15cm}
\subsection{Subspace Clustering Algorithms}\label{sec:algo}

The first component of the CoreSC package contains the actual subspace clustering algorithms.  
In our extension, the user can select among a multitude of different algorithms. These algorithms include grid based clustering techniques (CLIQUE \cite{DBLP:conf/sigmod/AgrawalGGR98}, DOC/FastDOC \cite{DBLP:conf/sigmod/ProcopiucJAM02}, MineClus \cite{mineclus}, SCHISM \cite{DBLP:conf/icdm/SequeiraZ04}), DBSCAN-based techniques (FIRES \cite{DBLP:conf/icdm/KriegelKRW05}, INSCY \cite{DBLP:conf/icdm/AssentKMS08}, SUBLCU \cite{DBLP:conf/sdm/KroegerKK04}) and optimization-based techniques for subspace clustering (PROCLUS~\cite{proclus}, 
STATPC \cite{DBLP:conf/kdd/MoiseS08}). 

Each algorithm implements the interface {\ttfamily SubspaceAlgorithm}, which defines the input and output of the algorithms. 
The input corresponds to a database of objects described by numerical features, i.e.\ each algorithm needs to be provided with a list of objects $\left\langle o_1,\ldots,o_n \right\rangle$ where $o_i \in \mathbb{R}^d$. 
The output of each algorithm is a list of subspace clusters $\left\langle C_1,\ldots,C_k\right\rangle$. Each subspace cluster $C_i$ represents the objects and relevant dimensions belonging to this clusters. Note that in subspace clustering each cluster has its individual set of relevant dimensions (cf.~Figure~\ref{fig:intro}). 
Thus, each subspace cluster corresponds to a tuple $C_i=(O_i,S_i)$ where $O_i$ represents the clustered objects by their objects ids, i.e.\ $O_i\subseteq \{1,\ldots,n\}$, and $S_i$ represents the relevant dimensions of the cluster, i.e.\ $S_i\subseteq \{1,\ldots, d\}$.

It is worth mentioning that subspace clustering in general is not restricted to disjoint clusters; thus, the result set might contain clusters $C_i$ and $C_j$ (with $i\neq j$) where $O_i\cap O_j\neq\emptyset$ or $S_i\cap S_j\neq\emptyset$. Additionally, dependent on the chosen algorithm, not necessarily each object or dimension needs to be part of some cluster, i.e.\ it might hold $\bigcup_{i=1}^k O_i\neq DB$ or $\bigcup_{i=1}^k S_i\neq \{1,\ldots,d\}$.

\begin{figure*}[t]
	\centering
		\includegraphics[width=1\textwidth]{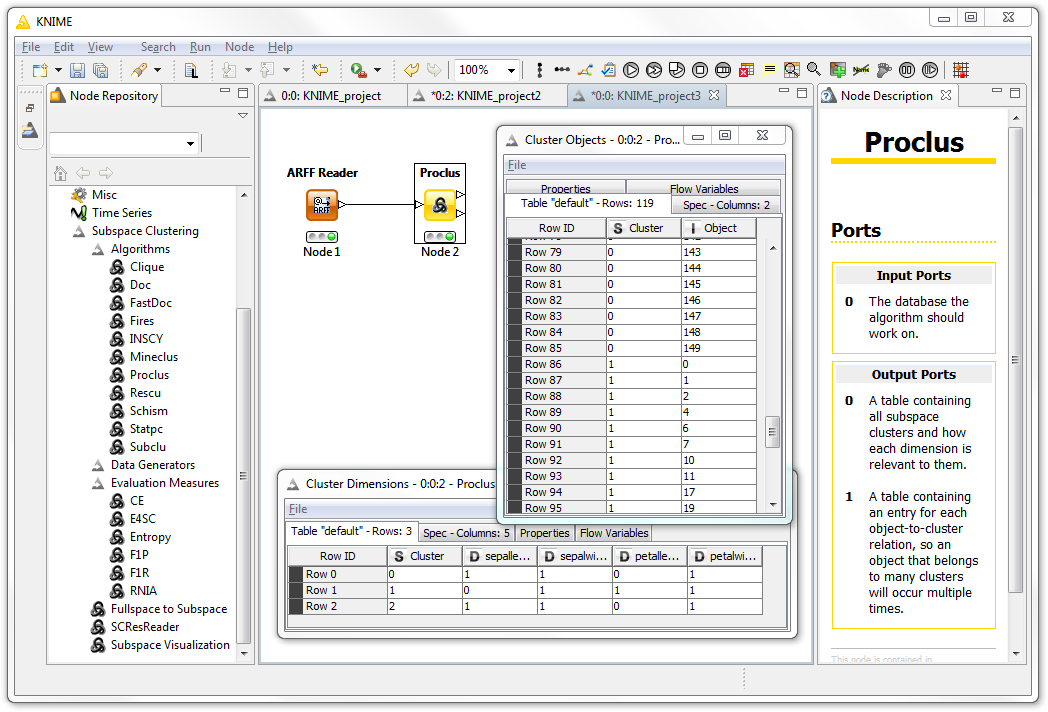}
	\caption{Screenshot of the subspace clustering extension for KNIME (left: newly developed nodes; center: workflow; right: description of nodes)}\label{fig:screenshot1}
\end{figure*}

\subsection{Data Generators}
The second component of the core package contains a flexible data generator first introduced in \cite{ref:KnimeSC}, which generates synthetic data with hidden subspace clusters. These datasets can be used to evaluate the correctness of subspace clustering algorithms and to assess the methods' scalability. The data generator implements the interface {\ttfamily SubspaceDataGenerator} which defines the two outputs of the generator.

The first output corresponds to the generated data, i.e.\ as above it corresponds to a list of objects $\left\langle o_1,\ldots,o_n \right\rangle$ with $o_i \in \mathbb{R}^d$. The second output of each data generator is the ground truth clustering. This ground truth specifies which clusters are hidden in the data and which clusters should be found by the subspace clustering algorithms. Accordingly, the second output is a list of subspace clusters $\left\langle C_1,\ldots,C_k\right\rangle$.

\subsection{Evaluation Measures}
The third component provides implementations of evaluation measures for subspace clustering. Evaluation measures summarize the clustering result by a numerical value where, e.g., a high value indicates better quality of the clustering.
Evaluation measures can be categorized into internal measures and external measures. While internal measures assess the quality of a clustering based on properties as, e.g., the compactness or density, external measures compute the quality w.r.t.\ a ground truth clustering \cite{farber2010using,E4SC}. 
Please note that the ground truth clustering can be any clustering: either generated by a data generator, provided manually by the user, or determined by an algorithm. Thus, besides comparing the result of a single algorithm against the ground truth, external measures can also be used to compare the results of two different algorithms on the same data. We provide several evaluation measure specifically designed for subspace clustering in our framework. These measure include CE, RNIA, Entropy, F1P, F1R, and E4SC. We kindly refer to \cite{E4SC} for a description of these measures.

In our extension, all evaluation measures implement the interface {\ttfamily SCEvaluationMeasure}. The interface specifies the input of these measure which corresponds to the database on which the clustering is performed, and two subspace clustering results $\left\langle C_1,\ldots,C_k\right\rangle$ and $\left\langle C'_1,\ldots,C'_l\right\rangle$. The output of each measure is a numerical value summarizing the quality of the clustering. Since some measures provide more fine grained evaluation results for each cluster individually, we additionally implemented the interface {\ttfamily SCExtendedEvaluationMeasure}. This interface allows to retrieve a evaluation result for each cluster of the result individually.

\newpage\subsection{Visualization}
The last component of the core package provides subspace clustering specific visualization and interaction principles. In our extension we integrated the CoDA \cite{gunnemann2010coda}, MCExplorer \cite{gunnemann2010mcexplorer}, and Visa \cite{assent2007visa} toolkits. While these techniques are independent of the used KDD framework, we additionally integrated further techniques exploiting the individual visualization capabilities of each framework. These methods are integrated in the framework-specific packages of KDD-SC.

\section{KNIME Extension}\label{sec:knime}
This section describes the usage of our extension within KNIME, termed KnimeSC.
We demonstrated a first version of KnimeSC at \cite{ref:KnimeSC}.
KNIME is an opensource data mining framework offering several benefits and is widely been used in industry as well as in academia.  
It has a modern, user-friendly interface which allows to model data mining workflows in an intuitive manner. 
In KNIME, a workflow is defined by a set of nodes, which can represent data sources and sinks, mining algorithms, transformations, visualizations, and further concepts.
Each node has specific input and output ports depending on the node's functionality. 
The user establishes a new workflow by selecting a set of nodes from the node repository and then connects the corresponding input and output ports 
to steer the data flow between these nodes. 
Data mining workflows can be stored for later re-use, modification, or extension. 

A major benefit of KNIME is the easy-to-use plugin concept. 
It allows KNIME to be extended by new features, represented as new nodes in the node repository. 
These new nodes can freely interact with the existing KNIME components, achieving a deep integration of our extension.
Thus, all techniques already integrated in KNIME can be combined with our extension for mutual benefit.

Figure~\ref{fig:screenshot1} shows a screenshot of KNIME and our extension.
On the left, the newly developed nodes are illustrated in the node repository.
Each node corresponds to one functionality provided by the CoreSC package.
 On the right, descriptions of each node and its corresponding input/output ports are given. In the center, the actual workflow is illustrated. In the following we provide details of our extension and the different types of nodes based on three different workflows.

\begin{figure}[t]
	\centering
		\includegraphics[width=0.41\textwidth]{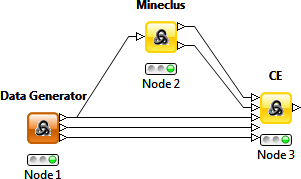}
	\caption{Workflow to evaluate the result of a subspace clustering algorithm w.r.t.\ a ground truth clustering (synthetically generated).}\label{fig:screenshot2}
\end{figure}

\subsection{Subspace Clustering Algorithms}
Figure~\ref{fig:screenshot1} shows a simple workflow where a data reader node ('Node 1'; here: reading data from an ARFF file) is connected with a subspace clustering algorithm node ('Node 2'; here: PROCLUS \cite{proclus}). Accordingly, by specifying this workflow, the user applies PROCLUS on a given database.

As described in Section~\ref{sec:algo}, each algorithm gets as an input the database to be clustered. This is shown by the single input port of the node 'Node 2'. Considering the output of the node, we have to take care of the special format used in KNIME. The standard format to exchange information between nodes in KNIME is by using flat tables/relations. Since the output of each algorithm is a list of subspace clusters, which itself are tuples describing sets of objects and sets of dimensions, each node needs to have two output ports. At the first output port, a table is provided which describes the relevant dimensions $S_i$ of each cluster $C_i=(O_i,S_i)$ via binary encoding. In Figure~\ref{fig:screenshot1} this table is illustrated at the bottom ('Cluster Dimensions') and shows three subspace clusters found in the Iris dataset \cite{UCISite}. The cluster with ID 2, for example, is located in the dimensions 'sepallength', 'sepalwidth', and 'petalwidth'.
The second output port provides information which objects belong to the detected clusters (table 'Cluster Objects'). In the example, the object 149 belongs to cluster 0, while object 17 belongs to cluster 1. Please note again that in subspace clustering each object might belong to \emph{multiple} clusters, i.e.\ clusters might overlap due to different subspace projections. Thus, the output table corresponds to an $n$:$m$ relation. These two outputs can be forwarded to any other node included in the KNIME framework as we will show next.

\begin{figure}[t]
	\centering
		
		\includegraphics[width=0.45\textwidth]{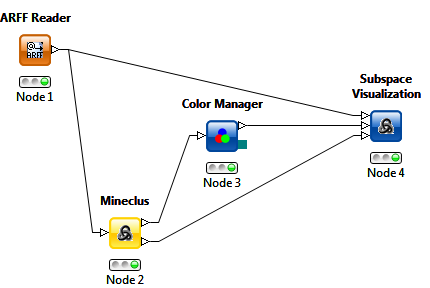}
		\includegraphics[width=0.45\textwidth]{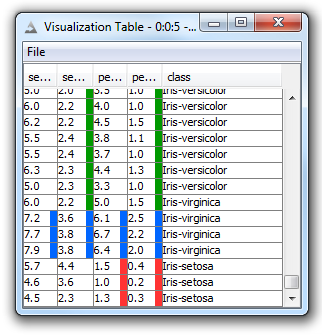}
	\vspace*{-1mm}
	\caption{Workflow to visualize the result of a subspace clustering algorithm via colored tables.}\label{fig:screenshot3}\vspace*{-2mm}
\end{figure}

\subsection{Data Generators \& Evaluation Measures}
A second workflow is illustrated in Figure~\ref{fig:screenshot2}. It models the task frequently performed in scientific literature: a) generate synthetic data with a given clustering ground truth, b) apply an algorithm on the data, and c) measure whether the detected result matches the ground truth.

\begin{figure*}[t]
	\centering
		\includegraphics[width=1\textwidth]{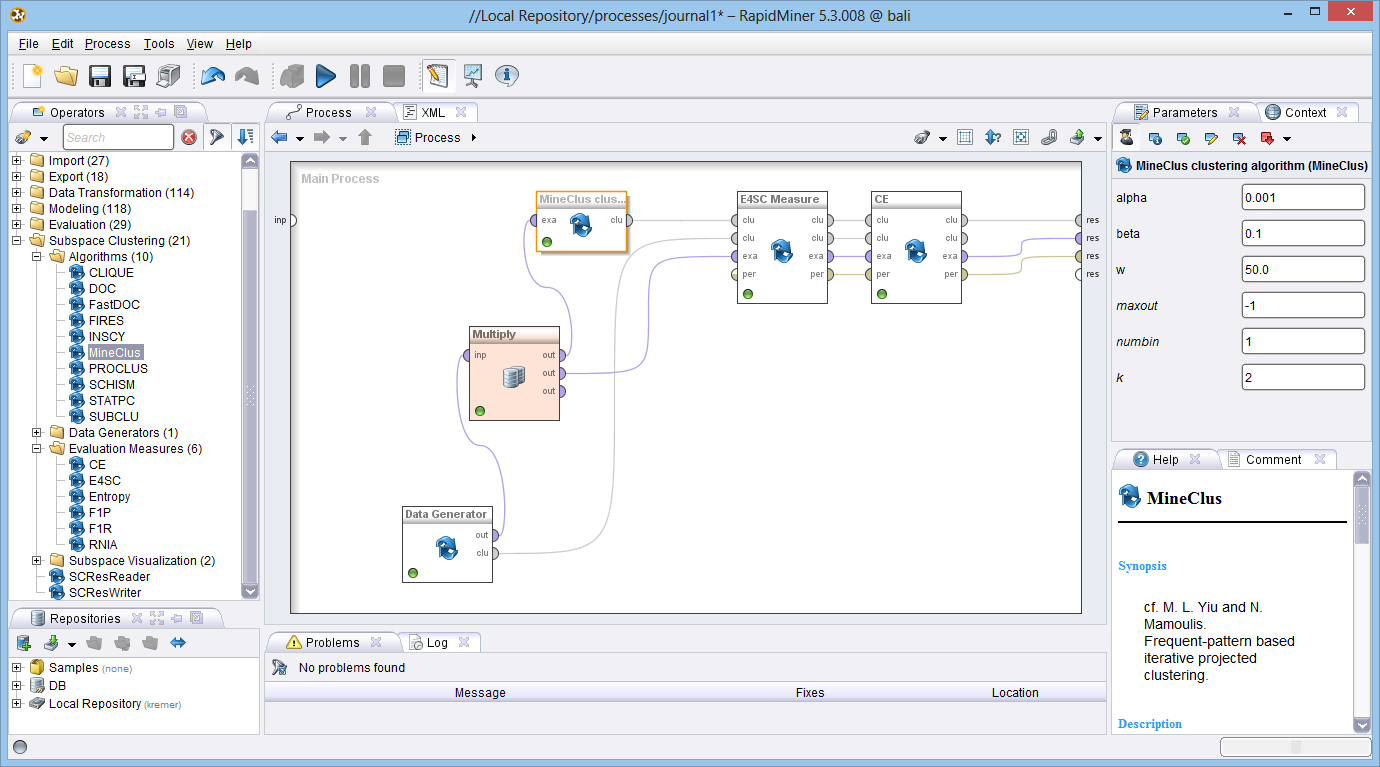}
	
	\caption{Screenshot of the subspace clustering extension for RapidMiner. (left: newly developed nodes; center: workflow with chained evaluation measures; right: parametrization and description of nodes)}\label{fig:RMscreen1}
	\vspace*{-2.5mm}
\end{figure*}

To solve this task with our framework, the user has to select a data generator node ('Node 1'). The node constructs synthetic data where the subspace clustering structure is known, i.e.\ the ground truth for clustering is given. Again, we have to take care that in KNIME the information is exchanged via flat tables. Thus, each data generator node has three output ports: First, the generated data. Second, the relevant dimensions of each cluster. Third, the cluster memberships of each object. The last two outputs are of the same format as the outputs of the subspace clustering algorithm nodes as described above. Connecting the first port of the data generator with an algorithm node ('Node 2') allows to cluster the synthetic data.

Finally, to measure the quality of the detected results, the user can use an evaluation measure node ('Node 3'; here: CE measure \cite{DBLP:journals/tkde/PatrikainenM06}). Such a node has five input ports: four ports are required to specify the two clustering results that should be compared (two ports for each clustering result), and one port for the database. Thus, in the figure, all three output ports of the data generator are connected to the measure as well as the two output ports of the MineClus node.

\subsection{Visualization}
The two outputs of each algorithm node already allow to analyze the detected clustering structure on a basic level. That is, by inspecting the corresponding tables (cf.\ Fig.~\ref{fig:screenshot1}) the user might get an impression about the relevant dimensions of the clusters and the supporting objects. Though, analyzing these tables might be difficult to gain further knowledge; accordingly, for easy interpretations of the clustering results we include different visualizations.

One possible visualization is realized with the workflow depicted in Figure~\ref{fig:screenshot3}. The subspace visualization node generates results as shown in the table on the right. The table represents the original database where each row corresponds to one object. The objects belonging to the same cluster are highlighted with the same color. In the example, three clusters are shown. Additionally, also the relevant dimensions of the clusters are depicted. A dimension is relevant, if and only if there is a colored bar on the right hand side of the number. In the example, the green cluster is located in the subspace of dimension 2 and 4, while the blue cluster is located in all four dimensions. Using this visualization, the user can easily compare the different subspaces of the clusters as well as the attribute values of the clustered objects. Considering for example the green cluster, we see that the attribute values in the first (and irrelevant) dimension are distributed in the broad range of 5.0-6.3, while the second (and relevant) dimension shows a deviation of only 2.0-2.4.

To obtain this visualization, the subspace visualization node requires three inputs: First, the database to be analyzed. Next, the relevant dimensions of each cluster with their corresponding coloring. This coloring is realized by using the Color Manager node provided by the KNIME framework. That is, the first output of the subspace clustering algorithm ('Node 2') is firstly forwarded to the Color Manager ('Node 3') before used as an input of the subspace visualization node ('Node 4'). In the Color Manager node, the user can choose the color of each cluster. The last input required for the visualization is the cluster membership information which can be directly transferred from 'Node 2'.

\section{RapidMiner Extension}\label{sec:rapid}

In this section we present the usage of our extension within the RapidMiner framework. Similar to KNIME, RapidMiner models data mining workflows via a node-based interface, i.e.\ each node performs a certain task and has specific input and output ports. Information between different nodes is exchanged by connecting their corresponding ports.

Figure~\ref{fig:RMscreen1} shows a screenshot of RapidMiner and our extension. On the left of the screen, the newly developed nodes are shown. On the right, the parametrization of the currently selected node is illustrated (here: the parameters of the MineClus algorithm) and a description of the node is provided. In the center, one sees the actual workflow.

While the general interaction with RapidMiner is similar to KNIME, we briefly discuss some differences. The exchange of information between KNIME nodes is based on flat tables. Thus, we represented a subspace clustering results via two flat tables, describing the object groupings and the relevant dimensions of the clusters. In RapidMiner, information between nodes is exchanged based on Java objects. Thus, instead of using flat tables, we directly exchange the list of subspace clusters via the Java class {\ttfamily SubspaceClusterModel}. Accordingly, in RapidMiner each node representing a subspace clustering algorithm has only a single output port (cf.~the MineClus node in Figure~\ref{fig:RMscreen1}) and each data generator node has only two outputs (one port for the ground truth clustering and the other port for the generated database). 
These output ports are typed, i.e.\ they can only be connected to other input ports which also accept subspace clustering results.

In Figure~\ref{fig:RMscreen1}, we see how the ground truth clustering of the data generator and the result of MineClus are forwarded to the evaluation measure E4SC. Additionally, the generated database is provided as an input for the measure.
The fourth input port shows a further feature integrated in RapidMiner: the chaining of nodes. Here, the measures E4SC and CE are chained, i.e.\ all output ports of E4SC act as input ports for CE. While the first three output ports simply forward the three input ports of the measure, the last output port represents the result of the evaluation measure and of all measures which are before this node in the chain. That is, in the workflow of Figure~\ref{fig:RMscreen1}, the result of the CE measure is a list representing the result of the E4SC and the CE measure. By chaining the nodes the workflows are more compact and, thus, easier understandable.      

 \subsection{Visualization}
\begin{figure}[t]
	\centering
		\includegraphics[width=0.475\textwidth]{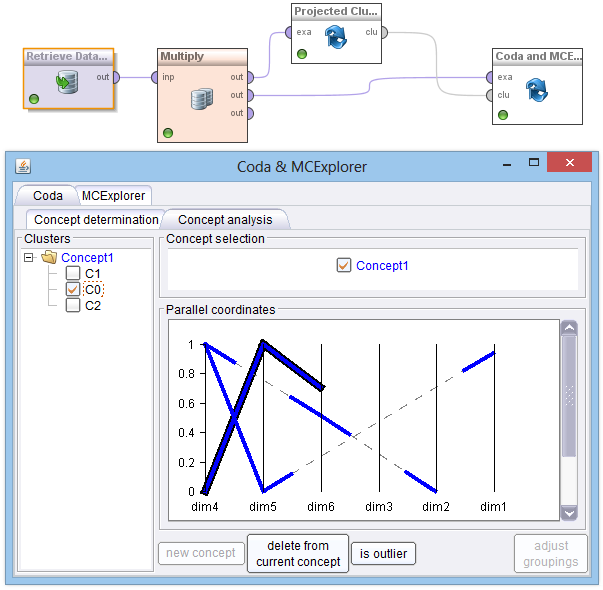}\vspace*{-0.5mm}
	\caption{RapidMiner workflow for applying the CoDA \& MCExplorer visual analysis of a subspace clustering result.}\label{fig:RMscreen2}\vspace*{-0.5mm}
\end{figure}

In Figure~\ref{fig:RMscreen2} we illustrate another RapidMiner workflow modeling the analysis of subspace clustering results via the CoDA \cite{gunnemann2010coda} and MCExplorer \cite{gunnemann2010mcexplorer} toolkits. After reading data from an external source, the database is forwarded to the PROCLUS node. The clustering result of PROCLUS and the database are then transferred to the visualization node shown on the right. Please note that in RapidMiner we have to use a so called 'Multiply' node when a single output port needs to be connected to multiple input ports. In the example, the loaded database is used as an input for PROCLUS as well as for the visualization. When activating the node on the right, a new window containing the CoDA and McExplorer toolkits will open in which the user can interact with the clustering result. A detailed description of the toolkits' functionalities is given in the original papers.

\section{WEKA Extension}\label{sec:weka}
\begin{figure*}[t]
	\centering
		\includegraphics[width=0.925\textwidth]{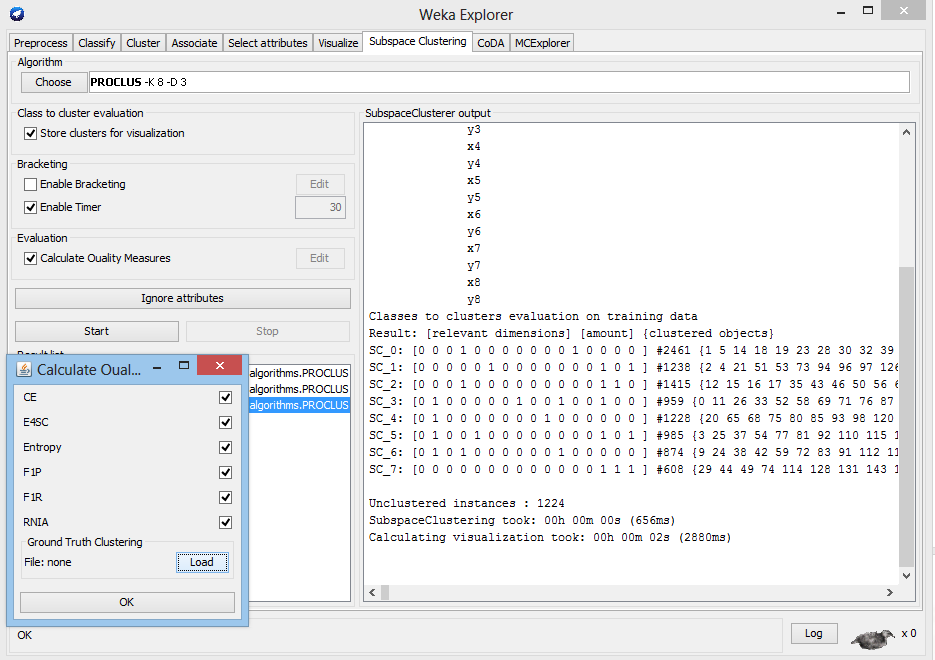}
	\vspace*{-1mm}
	\caption{Screenshot of the subspace clustering extension for WEKA. Shown are the results of a PROCLUS run and the possible Evaluation Measures that can be applied. }\label{fig:Wekascreen1}\vspace*{-0.5mm}
\end{figure*}

Finally, our extension is integrated into WEKA as shown in Figure~\ref{fig:Wekascreen1}. 
The functionality and user interface of WekaSC correspond to OpenSubspace \cite{DBLP:journals/pvldb/MullerGAS09}. 
As already mentioned, the advantage introduced by our redesign is its implementation of the WEKA plugin interface and its easy extensibility by the component-based design.

The classical workflow to analyze data in WEKA differs from the previous two knowledge discovery frameworks. It represents primarily a sequential process where a single dataset is loaded, preprocessed, and finally analyzed by an algorithm. The loading and preprocessing functionality of WEKA is integrated into the 'Preprocess' tab as shown in Figure~\ref{fig:Wekascreen1}.
When integrating our extension into WEKA, three novel tabs appear.

The 'Subspace Clustering' tab provides the major functionality of our extension. Here, the preprocessed data is analyzed using subspace clustering methods. The user can select among the multitude of implemented subspace clustering algorithms. In the example, the PROCLUS method is chosen. Additionally, the user can select different evaluation measures which will be applied when the result of the algorithm has been generated. As shown in the lower left part of the screenshot, different measures can be selected and the ground truth clustering to which the result is compared can be loaded.
After starting the algorithm, the clustering result will appear in the right part of the window. It represents the list of detected subspace clusters with their relevant dimensions in binary encoding as well as the number of objects per cluster and the corresponding object ids. This textual output corresponds to the two tables as used in KNIME.
The two remaining tabs 'CoDA' and 'MCExplorer' can be used to analyze the clustering results based on the corresponding toolkits as described before.

%% file: scenario.tex
\section{Conclusion}
Subspace clustering is an important mining task and it is widely studied in the scientific community. In this paper, we presented our subspace clustering extension KDD-SC, which is integrated in the KDD frameworks KNIME, RapidMiner, and WEKA. Our extension provides subspace clustering functionality for these frameworks based on a common code basis, and it can flexibly be combined with the toolkits' existing features.
Our KDD-SC extension is publicly available on the following website:
\vspace*{-1mm}\begin{verbatim}
http://dme.rwth-aachen.de/KDD-SC 
\end{verbatim}
\vspace*{-1mm}Overall, our extension sets the stage for the wide applicability of subspace clustering in practical applications.

\noindent {\textbf{\textit{Acknowledgment.}}} \hspace{-1.6mm} 
We thank Emmanuel Müller, Ira \text{Assent}, and Timm Jansen for their excellent work on the OpenSubspace project, which is a foundation of CoreSC and WekaSC.